\documentclass[nofootinbib,preprint,pra,showkeys]{revtex4-1}

\usepackage[final]{graphicx}% Include figure files
\usepackage{epstopdf}
\usepackage{verbatim}
\usepackage{textcomp}
\usepackage[toc,page]{appendix}
\usepackage{dcolumn}% Align table columns on decimal point
\usepackage{bm}% bold math
\usepackage{color}

\graphicspath{{./Fig/}}
\begin{document}
%\preprint{FV1.5}

\title{Dynamics of the ions in Liquid Argon Detectors and electron signal quenching}
\author{Luciano Romero} 
\author{Roberto Santorelli}\email[Corresponding author. E-mail address: ]{roberto.santorelli@ciemat.es.} 
\author{B\'arbara Montes }
\affiliation{Centro de Investigaciones Energ\'{e}ticas, Medioambientales y Tecnol\'{o}gicas (CIEMAT) Av. Complutense 40, 28040 Madrid, Spain}
%\collaboration{xxx}

\date{\today}

\begin{abstract}
A  study of the dynamics of the positive charges in liquid argon  has been carried out in the context of the future massive  time projection chambers proposed for dark matter and neutrino physics. Given  their small mobility coefficient in liquid argon, the ions spend a considerably longer time  in the active volume with respect to the electrons. The positive charge density can be additionally  increased by the injection, in the liquid volume, of the ions  produced by the electron multiplying devices located in gas argon. The impact of the ion current on the uniformity of the field has been evaluated as well as the probability of the charge signal  quenching due to the electron-ion recombination along the drift. The study results show some potential concerns for massive detectors with drift of many meters operated on surface. 
\end{abstract}

%\pacs{ PACS }% PACS, the Physics and Astronomy
                             % Classification Scheme.
                              
\maketitle
 
\tableofcontents

%%%%%%%%%%%%%%%%%%%%%%%%%%%%%%%%%%%%%%%%%%%%
%% MAINMATTER
%%%%%%%%%%%%%%%%%%%%%%%%%%%%%%%%%%%%%%%%%%%%

\section{Introduction} 

Liquid argon (LAr) detectors  have been widely used during  recent years in several fields ranging from neutrino physics \cite{Rubbia:2011ft, Anderson:2012mra, Ignarra:2011yq} to direct dark matter searches \cite{Benetti:2007cd, Badertscher:2013ygt, Agnes:2014bvk}, \textcolor{black}{given their  particle identification  and low energy threshold capabilities in large active volumes \cite{Chepel:2012sj}}. In particular a massive liquid argon time projection chamber (LAr-TPC) is the \textcolor{black}{chosen design}  for the next generation of  underground neutrino observatories recently proposed \textcolor{black}{\cite{DUNECDR}}. 

Particle interactions in  argon produce simultaneous excitation and ionization of the atoms,  generating VUV photons and ion/electron pairs.  In a typical LAr-TPC,  photon sensors are used to detect the  scintillation light, while a constant electric field $\vec{E_{d}}$ drifts the electrons to the anode. The charge readout can be carried out through the  collection of the electrons on thin wires placed directly in the liquid \cite{Amerio:2004ze} or, for double phase liquid-vapor detectors, through their extraction to a gas region placed above the sensitive volume \cite{WA105:technical}. In this case, a Townsend  avalanche can be induced through high electric fields, producing an amplified signal proportional to the number of primary electrons extracted from the liquid phase.  

\textcolor{black}{Both single and double phase options are presently investigated for the  DUNE experiment {\cite{DUNECDR}}, with maximum electron drifts of 3.6 m and 12 m respectively.} Other experiments foreseeing  drift  up to 20 m have been recently proposed  \cite{Murphy:2015uma, Cline:2006st, Angeli:2009zza, Agostino:2014qoa}, which require a considerable technological effort to mantain  a  level of contamination less than 60 ppt of O$_{2}$ equivalent  \cite{Tope2014} (electron half life $>$ 5 ms \cite{Buckley}),  in order to  reduce the impact of the electron quenching  by electronegative impurities contaminating the LAr bulk. A direct charge readout with the wires in a single phase chamber has the advantage of an overall simplified  detector design, \textcolor{black}{while} the amplification in the gas phase makes it possible to detect smaller charge signals, thus allowing it to reach a lower energy threshold, or \textcolor{black}{to exploit}  longer drift distance\textcolor{black}{s} with respect to the single phase design.  

The positive and negative charges, produced by the particle interactions in the liquid, drift to the  cathode and the anode following the same field lines, although the former have a drift speed which is six orders of magnitude lower than the latter ($v_{i}  \ll v_{e}$)   \cite{Walkowiak:2000wf,Dey:1968}. As a consequence,  the \textcolor{black}{positive} ions spend more time in the liquid  before they get collected on the cathode and neutralized, and the \textcolor{black}{ion} charge density is much larger than that of the electrons ($\rho_{i}  \gg \rho_{e}$). This effect can be particularly relevant for double phase detectors foreseeing large charge amplification factors, where the ions, created in the vapor volume, may drift back to the cathode crossing the gas-liquid interface  and \textcolor{black}{further} increase the $\rho_{i}$ in the \textcolor{black}{active volume}.
The space charge  can locally modify the amplitude of the electric field, the drift lines and the velocity of the  electrons  produced in the liquid, leading to a  displacement in the reconstructed position of the ionization signal. Additionally, the positive density $\rho_{i}$ can be sizable such that the probability of a ``secondary electron/ion recombination", different than the primary \textcolor{black}{electron/parent-ion} columnar recombination \textcolor{black}{\cite{Chepel:2012sj}}, has to be considered between the charge signal produced in the liquid and the ion current. The effect can cause an additional signal loss, with a probability dependent on the electron drift path, that could resemble the charge quenching given by the electronegative impurities in the active volume.

\textcolor{black}{In the present article we evaluated the impact of the positive charge density, produced by the cosmic rays and by the $^{39}\text{Ar}$ contamination in natural argon,  on the electron signal in massive detectors, evidencing, for the the first time, an intrinsic limit for the LAr technology given by the maximum  drift obtainable with a TPC operated with natural argon, even in case of a low background radiation environment and infinite liquid purity. This aspect can affect the dark matter experiments with a few meters drift only in case of very low drift fields ($\lesssim~100$~V/cm), however the study could be particularly relevant for the new generation of  neutrino experiments foreseeing drift paths of many meters, especially for what concerns the  supernovae neutrino  detection and low energy neutrino research.  Particularly, in Sec. \ref{Sec:res} we took into account the design of the protoDUNE detectors,  which will be operated on the surface, and the layout of the two single and double phase module types planned for the underground DUNE experiment. More detailed experimental studies and additional cases will be reported elsewhere \textcolor{black}{\cite{DMCIEMAT}}.}

\textcolor{black}{\section{Dynamics of the ions and impact of the interface gas/liquid}}

\begin{figure}[t!]
\begin{center}€
  \includegraphics[height=.35\textheight]{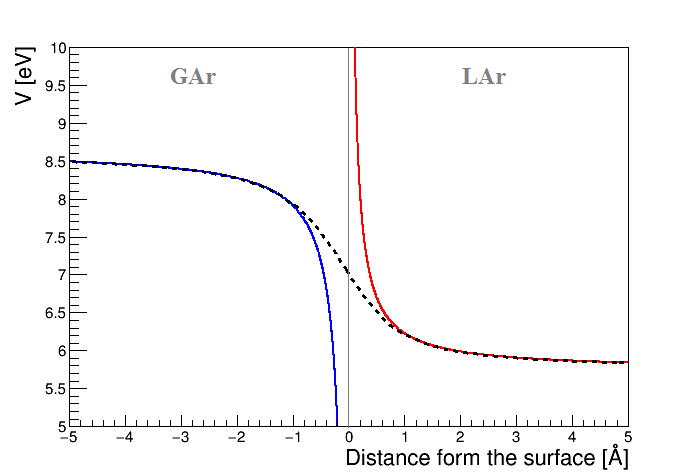}
  \caption{Potential energy (solid lines) at the liquid (right-red) / vapor (left-blue) interface in the mirror charge approximation  and possible effective potential energy (dashed line, see text for details).}
  \label{Pot_E_ion}
  \end{center}
\end{figure}

In a double phase detector, the \textcolor{black}{ionization electrons, produced by the particle interaction in the active volume,}  drift to the gas region where they are extracted and accelerated with the production of a Townsend avalanche. At the same time, given the low diffusion of the ions in gas argon relative to the typical size of the amplification region, a non-negligible fraction of the Ar ions produced by the avalanche can drift back to the liquid interface along the same field lines \textcolor{black}{followed by}  the extracted electrons.
When the distance between the ion and the liquid-vapor interface is greater than several angstroms, the liquid \textcolor{black}{is} treated as a continuum, thus an approximated description of the dynamics can be obtained solving a boundary condition problem between different dielectrics with the mirror charge method in a single dimension \cite{Griff}. Accordingly,  a point like charge $q$ in a medium with permittivity $\epsilon$, placed near the interface with another medium with permittivity $\epsilon'$, produces a mirror charge $q'=-q\cdot(\epsilon'-\epsilon)/(\epsilon'+\epsilon)$. Taking into account that the relative permittivity is $\epsilon_{LAr} = 1.5$ for liquid argon and $\epsilon_{GAr} = 1$ for argon vapor, the corresponding potential energy for an ion placed at a distance $d > 0$ from the liquid-vapor interface is a function of the inverse of the distance from the surface \cite{Bruschi:1975,Borghe:1990}:
\begin{eqnarray}
V_{LAr}(d)=\frac{q^2}{16\pi\epsilon_0\epsilon_{LAr}}\left(\frac{\epsilon_{LAr}-\epsilon_{GAr}}{\epsilon_{LAr}+\epsilon_{GAr}}\right)\frac{1}{d}+c_{LAr}\equiv\frac{A_{LAr}}{d}+c_{LAr}, \\
V_{GAr}(d)=\frac{q^2}{16\pi\epsilon_0\epsilon_{GAr}}\left(\frac{\epsilon_{GAr}-\epsilon_{LAr}}{\epsilon_{LAr}+\epsilon_{GAr}}\right)\frac{1}{d}+c_{GAr}\equiv\frac{A_{GAr}}{d}+c_{GAr},
\label{Potential_ion}
\end{eqnarray}

and it is depicted in Fig.~\ref{Pot_E_ion}, where the integration constants $c_{GAr}$ and $c_{LAr}$ account for the potential energy of the ion when it is far from the interface (see the Eq. \ref{Energy_ion}). 

Classically, the potential is infinite at $d = 0$, thus it has  been sometimes assumed that the barrier can preclude the ions from reaching the liquid phase  \cite{Bueno:2007um}, although that is true only if the charge can be approximated as point-like. Considering dimensions of the order of 1~\AA, as it is the case for the ionized atomic or molecular states whose formation is typical in noble gases \cite{Chepel:2012sj}, the mirror approximation is no longer valid. As the ion approaches the interface, it induces a displacement of the charge in the liquid that reduces the potential energy. The effective potential should decrease monotonically as the ion plunges into the liquid, following a sigmoidal shape (dashed line, Fig.~\ref{Pot_E_ion}), thus the problem is reduced to a finite classical potential barrier.

At the same time, the crossing of the liquid-gas interface is energetically favored. Considering the ion as a uniformly charged sphere of radius $a$, its potential energy far from the surface can be expressed as:
\begin{equation}
V=\frac{3}{5}\frac{q^2}{4\pi\epsilon a}.
\label{Energy_ion}
\end{equation}

Taking into account  that $a$ is of the order of $\approx1$~\textrm{\AA}, the $\approx2.9$~eV difference between the  potential energies at the interface  allows the injection of the ions into the liquid\footnote{According to the model presented, the difference between the potential energy of the ions at the liquid/gas interface  is one order of magnitude larger than that of the electrons (0.21 eV \cite{Bueno:2007um}).}, thus the possibility that a large fraction of the positive charge  produced in the gas phase enters the liquid cannot be discarded. 

For the present study it is irrelevant what  kind of charge amplification device is used: we introduce the ion gain  $G_I$ defined as the number of positive ions  injected into the liquid  for each electron extracted. This factor is proportional to the  electron  amplification $G$ through a constant $\beta$ ($\beta < 1$)  which takes into account the average loss of the positive charge in the gas, given by the ions scattering onto field lines not ending on the liquid surface\footnote{The value of $\beta$ depends on the geometry and the field configuration of the \textcolor{black}{specific} charge amplifying device. }, as well as the efficiency to pass the liquid-gas interface. If the amplification factor $G_I$ is large enough, the positive charge density $\rho_i$ in the liquid can be widely increased by the secondary ions produced by the avalanche.
 
\textcolor{black}{\section{Drift field distortion and electron-ion recombination}}

\begin{figure}[t!]
\begin{center}
  \includegraphics[height=.245\textheight]{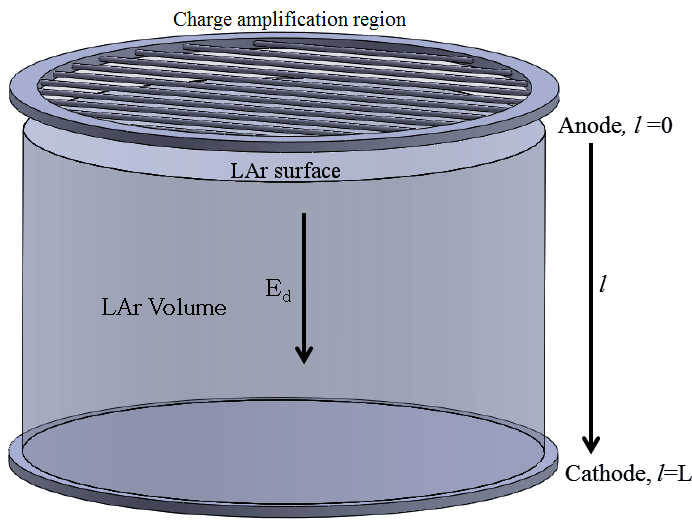}
  \includegraphics[height=.245\textheight]{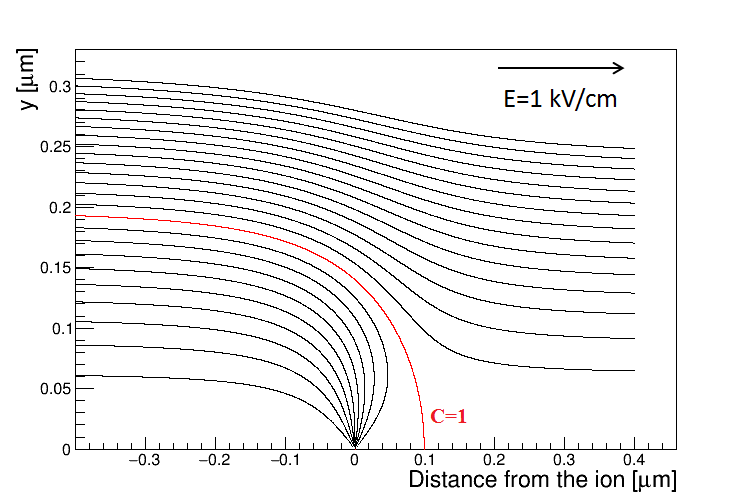}
  \caption{\textcolor{black}{Sketch of a typical double phase detector, with a charge amplification stage, considered in the present study ($left$). In case of a single phase TPC the readout with wires is at the anode of the drift field.} Configuration of the drift lines near an ion for 1~kV/cm \textcolor{black}{($right$)}. The red line, obtained for $C = 1$ (see appendix \ref{FieldL}), is the envelope of the path ending on the ion. }
  \label{Field_lines}
  \end{center}
\end{figure}

We consider a LAr-TPC with an axial geometry and the $l$ axis perpendicular to the surface of the liquid (\textcolor{black}{Fig.~\ref{Field_lines}$-left$}). The drift field $\vec{E_{d}}$ in the liquid  is along $l$, with the anode at the origin ($l=0$) and the cathode at a positive distance  $L$.
 In the following calculation we assume that the detector is wide enough such that the  transverse coordinate is not relevant for the discussion and $\vec{E_{d}}$ is constant in any transverse section of the detector. 

In the limit of a null ion current, the drift field is constant and it is equal to the cathode voltage divided by the total drift length $L$. On the contrary, \textcolor{black}{an} ion cloud makes  the drift field to change with $l$, \textcolor{black}{from a} minimum at the anode \textcolor{black}{to a}  maximum at the cathode. \textcolor{black}{In order to achieve the desired field value at the anode, it is necessary to increase the cathode voltage by a factor obtained integrating the drift field expression as a function of $l$ between 0 and $L$.}

Additionally, the possibility has to be considered  that an  ion recombines along the drift with a \textcolor{black}{ionization electron},  causing a quenching of the charge signal similar  to the one given by the electronegative impurities.  In order to  evaluate the  probability of that ``secondary" recombination, \textcolor{black}{which affects the free electrons and ions escaping} the primary columnar recombination \textcolor{black}{\cite{Chepel:2012sj}}, we define the cross section $S_{CS}$ as the transverse area whose crossing field lines end on one ion. The section should be far enough from the ion such that the ion field is negligible compared to the drift field, \textcolor{black}{and} all the lines emerging from the ion cross that section. Fig.~\ref{Field_lines}\textcolor{black}{$-right$} shows the field paths approaching an ion positioned at (0,0), which has a negligible size at the micron scale, in case of $E_d = 1$~kV/cm (see appendix~\ref{FieldL} for the detailed calculation). The red curve (C = 1) is the envelope of all the  lines ending on the ion.

The total number of field lines emerging from the ion, $q/\epsilon$, is equal to the number of lines traversing the cross section, $E_{d}\cdot S_{CS}$, therefore:
\begin{equation}
S_{CS}=\frac{q}{\epsilon E_{d}},
\label{Scs}
\end{equation}
where $q$ is the elementary charge, $\epsilon$ the absolute permittivity of the liquid argon and $E_d$ the amplitude of the drift field. For a typical $E_d$ value of  1~kV/cm,  $S_{CS}=1.2\cdot 10^{-7}~\text{mm}^2$  is much larger than the ion dimensions, thus the \textcolor{black}{total cross section,} obtained considering \textcolor{black}{all the ions constantly present within the drift volume  ($\approx~10^9-10^{11}$~ions/m$^3$ in average without considering the possible ion feedback from the gas, see Sec. \ref{Sec:res} and Eq.~\ref{Worst_probability})}, can be macroscopic and  the effect of the recombination cannot be neglected. 

We define the ion and electron fluxes as follows:
\begin{equation}
j_i(l)=v_i(l)\rho_i(l),\qquad j_e(l)=v_e(l)\rho_e(l),
\label{densities}
\end{equation}

where $\rho(l)$ and $v(l)$ are the  particle density and  velocity respectively at distance $l$, \textcolor{black}{the latter one being}  related  to the drift field through the mobility coefficient $\mu$ in liquid argon:
\begin{equation}
v_i(l)=\mu_i E_d(l),\qquad v_e(l)=\mu_e E_d(l).
\label{mobilities}
\end{equation}
Considering a typical drift field of $E_d = 1$~kV/cm, the experimental value for the electron velocity is $v_e \approx 2$~mm/$\mu$s \cite{Walkowiak:2000wf}. On the other hand, the  ion mobility, measured in steady state by subtracting the liquid motion, is $\mu_i \approx 2\cdot 10^{-4}~ \text{cm}^2\,\text{V}^{-1}\text{s}^{-1}$ \cite{Dey:1968}. Even considering the  more conservative values of $\mu_i \approx 1.6\cdot 10^{-3}~\text{cm}^2\,\text{V}^{-1}\text{s}^{-1}$ \cite{ICARUS:2015torti} the expected ion velocity is $v_i \approx 1.6\cdot 10^{-5}$~mm/$\mu$s, which is five orders of magnitude lower than that of the electrons.

The recombination rate, $r(l)$, in m$^{-3}$s$^{-1}$, is given by the ion density multiplied by the flux of the electrons and by the cross section:
\begin{equation}
r(l)=\rho_i(l)\, j_e(l)\, S_{CS}(l).
\label{rec_rate_1}
\end{equation}
Substituting $\rho_i$, $v_i$ and $S_{CS}$ from the Eq.~\ref{densities}, \ref{mobilities} and \ref{Scs} we obtain:
\begin{equation}
r(l)=\frac{j_i(l)\,j_e(l)}{v_i(l)}\,\frac{q}{\epsilon E_{d}(l)}=j_i(l)\,j_e(l)\,\frac{q}{\mu_i \epsilon E_{d}^2(l)}.
\label{rec_rate_2}
\end{equation}
\textcolor{black}{The last equation}  allows the determination of the charge signal loss in the liquid knowing the currents and the drift field, \textcolor{black}{whose expressions will be calculated in the following}.

\subsection{Particle currents and drift field equations}
The environmental radioactivity, the  $^{39}\text{Ar}$ decays and the cosmic muons continuously  produce ion-electron pairs  within the liquid argon active volume.   These sources are assumed to be uniformly distributed within the target, so we can introduce a constant ionization rate, $h$, defined as the average number of free pairs  escaping the primary recombination, per unit of time and volume.

In a stationary state, the ion and electron density variation should be null at any position, therefore:
\begin{equation}
0=h-r(l)-\frac{\text{d}j_i(l)}{\text{d}l},\qquad0=h-r(l)+\frac{\text{d}j_e(l)}{\text{d}l}.
\label{Zero_eq}
\end{equation}
The electron current \textcolor{black}{is maximum at the anode} and it diminishes with the axial distance, $l$, therefore the quantity $\text{d}j_e/\text{d}l$ is negative. \textcolor{black}{On the contrary the ion current increases with $l$ and it is maximum at the cathode.} Replacing  the recombination rate given by expression~\ref{rec_rate_2} in the Eq.~\ref{Zero_eq},  we get:
\begin{equation}
\frac{\text{d}j_i(l)}{\text{d}l}+j_i(l)\,j_e(l)\,\frac{q}{\mu_i \epsilon E_{d}^2(l)}=h,\qquad
\frac{\text{d}j_e(l)}{\text{d}l}-j_i(l)\,j_e(l)\,\frac{q}{\mu_i \epsilon E_{d}^2(l)}=-h.
\label{Dif_eq_1}
\end{equation}
On the other hand, the variation of the drift field, that we assume parallel to the detector axis, is determined by the charge density:
\begin{equation}
-q\,\rho_e(l)+q\,\rho_i(l)=\epsilon\frac{\text{d}E_d(l)}{\text{d}l}.
\label{Eq_campo_1}
\end{equation}
Since the ions spend considerably longer time in the liquid  before they get collected on the cathode, the positive charge density is much larger than that of the electrons. Being $\rho_{i}  \gg \rho_{e}$, we can disregard $\rho_e$ in the Eq. \ref{Eq_campo_1}. Considering the expression of the ion density from the Eq.~\ref{densities}, and using expression~\ref{mobilities}, we have:
\begin{equation}
j_i(l)=\frac{\epsilon\,v_i(l)}{q}\,\frac{\text{d}E_d(l)}{\text{d}l}=\frac{\epsilon\,\mu_i\,E_d(l)}{q}\,\frac{\text{d}E_d(l)}{\text{d}l}=\frac{1}{2}\frac{\epsilon\,\mu_i}{q}\,\frac{\text{d}E_d^2(l)}{\text{d}l}.
\label{Eq_campo_2}
\end{equation}
Equations~\ref{Dif_eq_1} and \ref{Eq_campo_2} are three coupled differential equations with three functions ($j_i$, $j_e$ and $E_d$) and one variable $l$. The argument of those equations can be simplified introducing the function
\begin{equation}
f(l)=\frac{\epsilon\,\mu_i}{q}\,E_d^2(l),
\label{Cambio_variable}
\end{equation}
then the 3 coupled linear equations stand:
\begin{equation}
\frac{\text{d}j_i(l)}{\text{d}l}+\frac{j_i(l)\,j_e(l)}{f(l)}=h,\qquad
\frac{\text{d}j_e(l)}{\text{d}l}-\frac{j_i(l)\,j_e(l)}{f(l)}=-h,\qquad
j_i(l)=\frac{1}{2}\,\frac{\text{d}f(l)}{\text{d}l}.
\label{Dif_eq_2}
\end{equation}
The three boundary conditions, which allow a particular solution of the equations, can be obtained taking  the electric field at the anode as parameter, and considering that the  electron current at the cathode is null and  the ion current at the anode  is given by the electron current multiplied by the ion gain:
\begin{equation}
E_d(0)=E_{A}.
\label{Emin}
\end{equation}
\begin{equation}
j_e(L)=0,
\label{Je_anode_null}
\end{equation}
\begin{equation}
j_i(0)=G_I\,j_e(0), 
\label{Ji_cathode}
\end{equation}

The system of equations given by \ref{Dif_eq_2}  can be solved numerically for any particular detector. In order to get an approximated solution we consider as negligible the impact of the recombination in the ion current equation. That is the case since  $\rho_{i}  \gg \rho_{e}$ even at low  or null $G_I$ values, given
the several  orders of magnitude lower drift speed of the ions with respect to the electrons. Accordingly  the recombination is disregarded  from the first equation of \ref{Dif_eq_2}:
\begin{equation}
\frac{\text{d}\,j_i(l)}{\text{d}l}=h\quad\rightarrow\quad j_i(l)=h\,l+j_i(0),
\label{Curr}
\end{equation}
however it has to be taken into account for the electrons, so the second equation of \ref{Dif_eq_2} still stands.
From the third equation of \ref{Dif_eq_2} and using equations \ref{Cambio_variable}, \ref{Emin} and \ref{Curr}  we have:
\begin{equation}
f(l)=2\int_0^l j_i(l)\,\text{d}l=h\,l^2+2\,j_i(0)\,l+f_0,\qquad f_0=f(0)=\frac{\epsilon\mu_i}{q}{E_{A}}^2.
\label{feq}
\end{equation}
Thus, taking into account the Eq. \ref{Cambio_variable}, the drift field as a function of $l$ can be written as:
\begin{equation}
E_d(l)=\sqrt{\frac{q}{\epsilon\mu_i}(h\,l^2+2\,j_i(0)\,l)+{E_{A}}^2},
\label{Drift_field}
\end{equation}
and it can be calculated  knowing \textcolor{black}{ the ion gain and the minimum field, which are boundary conditions, and the electron current at the anode, which has to be determined. In order to do that} the  second equation of \ref{Dif_eq_2} can be written using  \ref{Curr}  and  \ref{feq} as:
\begin{equation}
\frac{\text{d}j_e(l)}{\text{d}l}-\frac{(h\,l+j_i(0))}{h\,l^2+2\,j_i(0)\,l+f_0}j_e(l)=-h.
\label{Electron_current_dif_eq}
\end{equation}
Considering the boundary condition of a null electron current at the cathode (Eq. \ref{Je_anode_null}), the linear differential equation can be solved analytically:
\begin{equation}
j_e(l)=-h\,F(l)\,\ln \left( \frac{l+j_i(0)/h+F(l)}{L+j_i(0)/h+F(L)}\right),
\label{Electron_current_solution}
\end{equation}
where
\begin{equation}
F(l)=\sqrt{l^2+\frac{2\,j_i(0)}{h}\,l+\frac{f_0}{h}}.
\label{R_function}
\end{equation}

The electron current at the anode can be obtained calculating numerically $j_i(0)$ with the Eq. \ref{Electron_current_solution} taking into account the Eq. \ref{Ji_cathode}.

\section{Results and discussion} 
\label{Sec:res}

The  electron quenching probability and the effective drift field in a LAr-TPC \textcolor{black}{will} be calculated \textcolor{black}{in the following,} considering the average ionization produced in the argon target  by the $^{39}$Ar  decay and by the cosmic muons. Other contributions given,  for example, by the natural radioactivity, the Rn decay  or the material contamination, are considered negligible \textcolor{black}{in the present study}. 

In an underground facility the dominant contribution to the charge production is typically given by the  $^{39}$Ar decay, a $\beta$ emitter with a Q-value of 565~keV whose activity in natural argon  is $\approx1$~Bq/kg \cite{Benetti:2006az}, or, for liquid argon, $\approx1400$~Bq/m$^3$.
We can assume that the mean energy deposited  per decay in the active volume is approximately one third of the total Q-value. Considering that the commonly
 accepted value for the average energy required to create an ion-electron pair in LAr is W = 23.6~eV \cite{Miyajima:1974}, one $^{39}$Ar decay produces in average $\approx8\cdot10^3$~pairs, therefore the  ionization rate $h_0$  due to $^{39}$Ar is $\approx1.1\cdot10^7$~pairs/(m$^3$s).

\begin{figure}
 \begin{center}
  \includegraphics[height=.23\textheight]{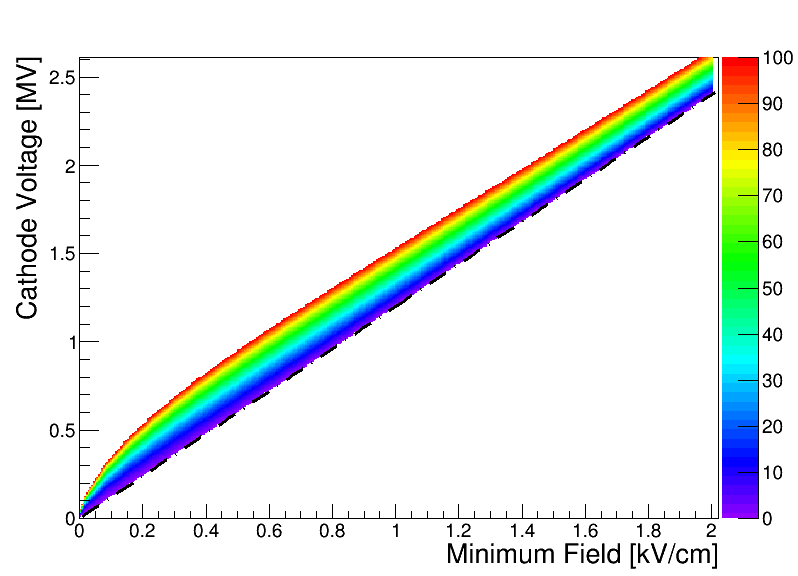}
  \includegraphics[height=.23\textheight]{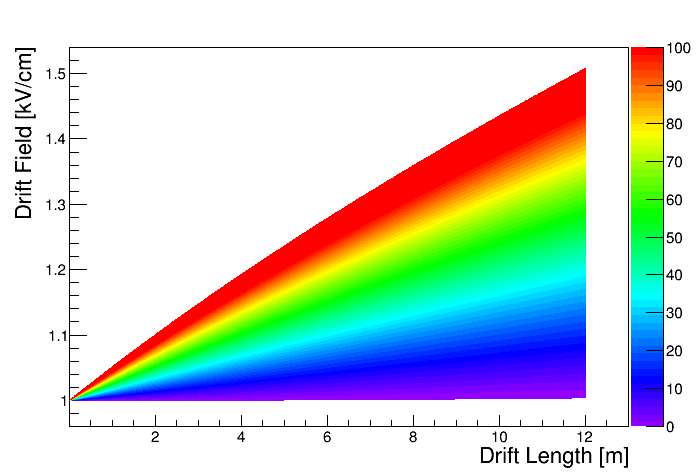}
  \caption{Underground case: cathode voltage as a function of the minimum drift field $E_A$, for different ion gains (color scale) in case of a   dual phase 12~m LAr detector ($left$). The dashed black line represents the voltage value required for a given  field in case of a null ion density. Variation of the field amplitude inside the same detector for different ion gains,  assuming a field at the anode of 1~kV/cm ($right$).}
  \label{HV_UG}
  \end{center}
\end{figure}

The prompt \textcolor{black}{columnar} recombination of the electrons with the parent ions  is a function of  the drift field and it  is usually approximated with the so-called Birks  law\footnote{We conservatively considered the minimum drift field or field at the anode  $E_{A}=E_{d}$.} $h=\frac{h_0}{1+k_E/ E_d}$ \cite{Birks:1964}, which gives the average free charge constantly produced in LAr by the $^{39}$Ar decay (see the Eq.~\ref{Zero_eq}). The constant $k_E$ has been experimentally measured in argon and it is equal to $0.53 \pm 0.04$~kV/cm~\cite{Scalettar:1982}.

In case the detector is located on the surface, the contribution to the total ionization produced in the active volume is mainly given by the muons, whose flux at sea level is reported to be $168~\text{muons}/(\text{m}^2\text{s})$ \cite{Grieder:2001ct}. Considering that most of the muons are at their minimum ionizing energy, the energy loss in argon as a function of the density is:
\begin{equation}
\frac{\text{d}E}{\text{d}l}\approx1.5\,\frac{\text{MeV}\,\text{cm}^2}{\text{g}},
\end{equation}

which gives an average deposited energy  $\text{d}E/\text{d}l=210$~MeV/m per muon or $35$~GeV/(m$^3$s) in liquid argon. Assuming the same W-value as before, the ion production rate is  $h_0=1.5\cdot 10^9$ pairs/(m$^3$s), two orders of magnitude bigger than that of the $^{39}$Ar decay.

The field variation within the detector and the cathode voltage needed to produce a given drift field can be evaluated by the integration of the Eq.~\ref{Drift_field} between 0 and $L$ (see appendix~\ref{KVolt}).  The  positive charge density makes the drift field to change with $l$  such that it is minimum at the anode and maximum at the cathode. In order to get the required minimum field,   the nominal cathode voltage,  calculated without taking into account the positive charge density, has to be increased by a factor dependent on the ion gain. Fig.~\ref{HV_UG}-$left$ shows the results in case of 12~m drift length with a detector placed in an underground laboratory, thus taking into account only the ionization  produced by the $^{39}$Ar decays in natural Ar. Important differences between the effective and the nominal field are expected at higher ion amplification values.  
The amplitude of the field within a detector is shown in Fig.~\ref{HV_UG}$-right$ as a function of the drift length  considering $E_A$~=~1~kV/cm and L~=~12~m. Depending on the ion gain, up to a 50\% difference is expected between the value at the anode and the one at the cathode, while variations of the order of 1~-~2\%  should be considered without the charge amplification   on such drift length.  

\begin{figure}
 \begin{center}
  \includegraphics[height=.23\textheight]{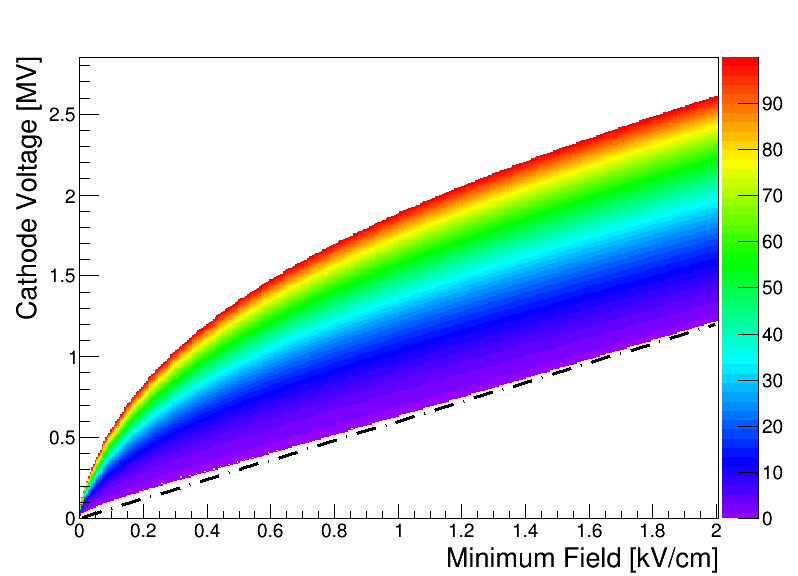}
  \includegraphics[height=.23\textheight]{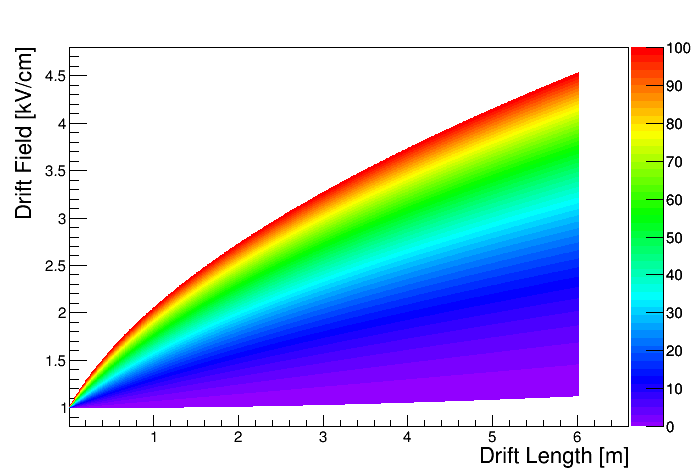}
  \caption{Surface case: cathode voltage as a function of the minimum drift field $E_A$, for different ion gains (color scale) in case of a dual phase 6~m LAr detector ($left$).  \textcolor{black}{The dashed black line represents the voltage value required for a given  field in case of a null ion density}. Variation of the field amplitude inside the same detector for different ion gains, assuming a field at the anode of 1~kV/cm ($right$).}
  \label{HV_Sup}
  \end{center}
\end{figure}

Fig.~\ref{HV_Sup} shows similar plots for a 6~m detector placed on the surface.   Given the two orders of magnitude larger ionization rate $h_0$ produced by the muons with respect to the $^{39}$Ar decays, the cathode voltage has to be slightly increased, even without charge amplification, in order to get the required \textcolor{black}{minimum} field amplitude. Cathode voltages of the order of one megavolt are required in order to be able to produce a field of 1~kV/cm \textcolor{black}{at the anode} for $G_I\approx$~10, while, for ion gains of the order of 20, twice the voltage is needed with respect to the nominal value calculated with a null positive charge density. The field variation inside the detector is shown in Fig. \ref{HV_Sup}$-right$.  \textcolor{black}{Cathode fields   with twice the amplitude at the anode  are necessary  in order to produce the required minimum field} for $G_I > 10$, evidencing the important non-uniformity of the field inside the detector. Even in case of $G_I = 0$, \textcolor{black}{as it is the case of a single phase detector,} differences of the order of 10\% are expected between the field at the anode and  the cathode because of the ion density produced by the cosmic rays in the liquid. \textcolor{black}{Given the significantly larger  ionization rate $h$ with respect to the underground case, the null ion density case (Fig.~\ref{HV_Sup}-dashed line) is not a realistic scenario.}

The impact of the electron-ion secondary recombination on the charge quenching can be evaluated considering the probability $P(l)$ that a free electron, created at depth $l$, reaches the anode. That probability is equal to the fraction of the surface S($l$) spanned by the field lines ending on the anode with respect to the total \textcolor{black}{anode} area $S(0)$, and it  is minimum for $l=L$ and one for $l=0$. Since the number of field lines ($E\cdot S$) is conserved all along the detector depth,  $E(0)\cdot S(0)=E(l)\cdot S(l)$, we obtain from the Eq. \ref{Drift_field}:
\begin{equation}
P(l)=\frac{S(l)}{S(0)}=\frac{E(0)}{E(l)}=\frac{E_{A}}{\sqrt{\frac{q}{\epsilon\mu_i}\,(hl^2+2j_i(0)l)+E_{A}^2}},
\label{Worst_probability}
\end{equation}

whose solution can be calculated knowing the constant ionization rate $h$, the field at the anode $E_{A}$ and the ion gain $G_I$, \textcolor{black}{which are input parameters}.

\begin{figure}[t!]
\begin{center}
 \includegraphics[height=.25\textheight]{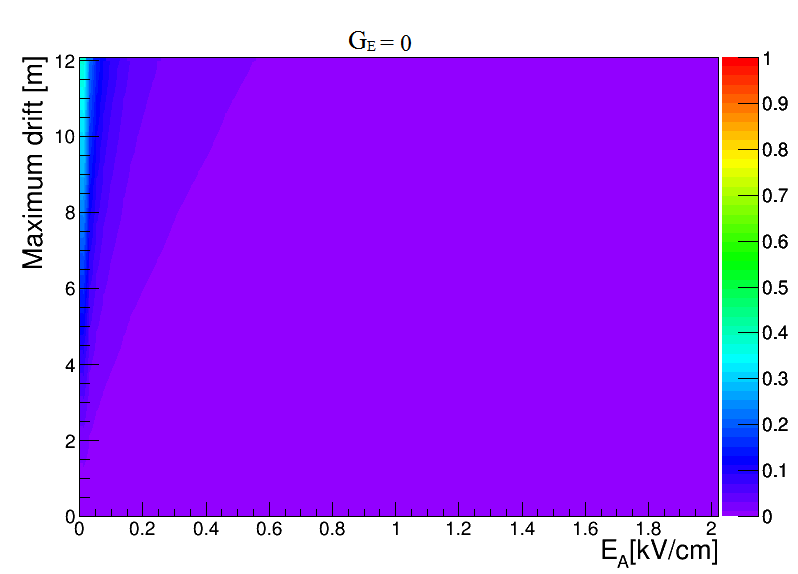}
 \includegraphics[height=.25\textheight]{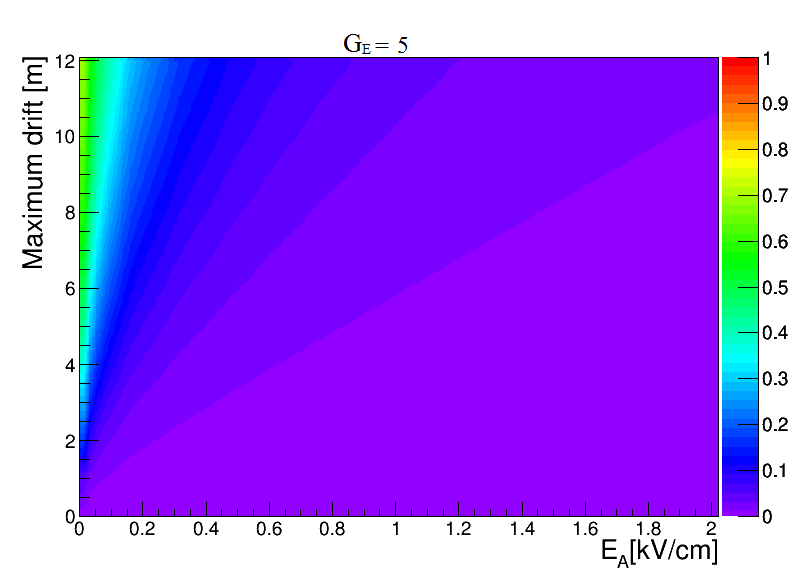}
  \includegraphics[height=.25\textheight]{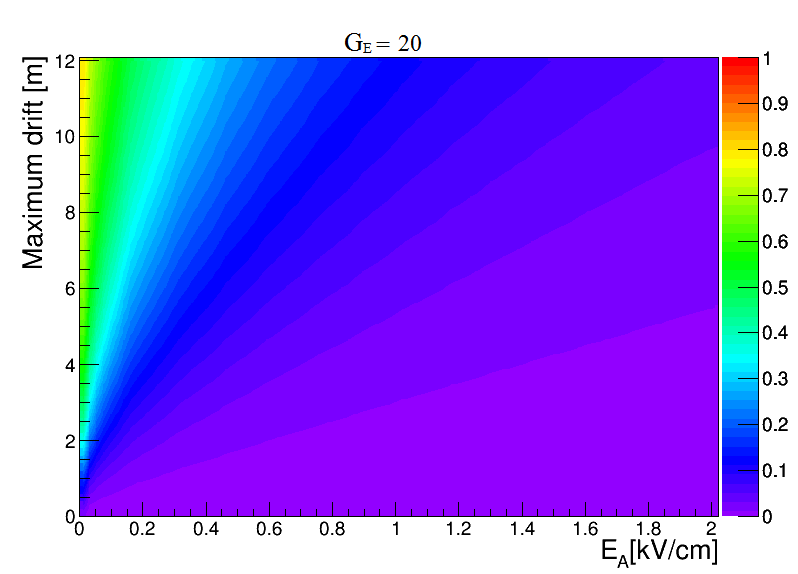}
   \includegraphics[height=.25\textheight]{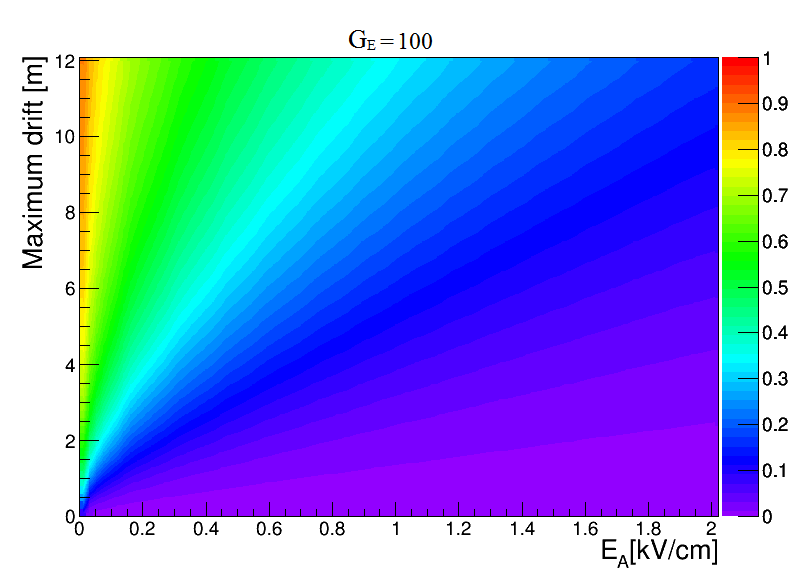}
  \caption{Underground case: recombination probability  (color scale) for electrons generated at the cathode level as a function of the maximum drift length L and the anode field E$_{A}$, considering  $G_I = 0$ ({\it{top-left}}), $G_I = 5$ ({\it{top-right}}),  $G_I = 20$ ({\it{bottom-left}}) and $G_I = 100$ ({\it{bottom-right}}).}
 \label{Losses_Underground_Gfixed}
  \end{center}
\end{figure}

The Eq.~\ref{Worst_probability} has been solved for a detector placed underground,  using the conservative value for the ion mobility ($\mu_i \approx 1.6\cdot 10^{-3}~\text{cm}^2\,\text{V}^{-1}\text{s}^{-1}$ \cite{ICARUS:2015torti}) and considering drift lengths up to 12~m and anode fields up to 2~kV/cm. The maximum charge quenching, evaluated considering the recombination probability of the electrons produced at distance $L$  from the anode, is  plotted in Fig.~\ref{Losses_Underground_Gfixed}  as a function of the drift length and field, assuming $G_I = 0$, $G_I = 5$, $G_I = 20$ and $G_I = 100$. Without the charge amplification, the recombination is practically negligible unless very low fields ($\lesssim~0.3$~kV/cm) and long drift distances ($\approx$~10~m) are foreseen.  At a typical field of $\approx$~1~kV/cm  we can expect relatively small charge signal losses ($\leq$ 10\%) for a few meters drifts and ion gains of the order of some tens, at the same time,  if $G_I$ is of the order of 100, more than 20\% of the charge signal created at the cathode level is quenched after 6~m drift and more than 50\% after 12~m.

\begin{figure}[t!]
\begin{center}
\includegraphics[height=.25\textheight]{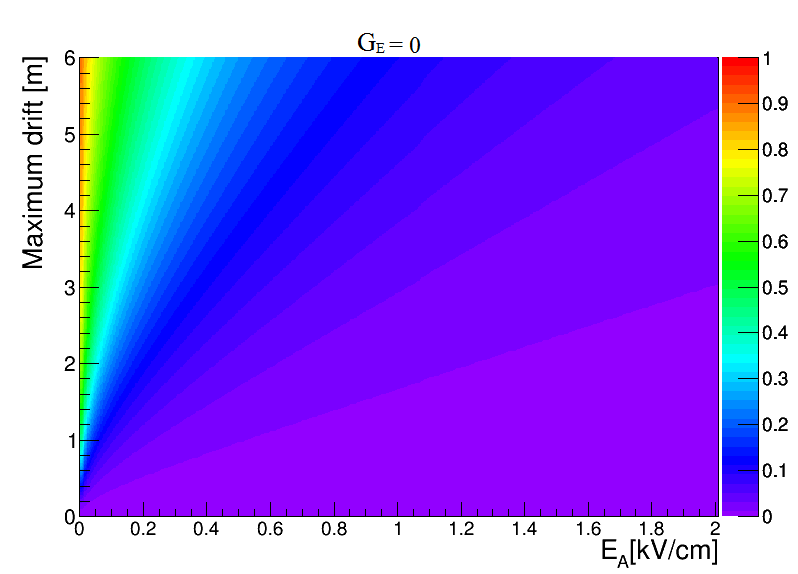}	
  \includegraphics[height=.25\textheight]{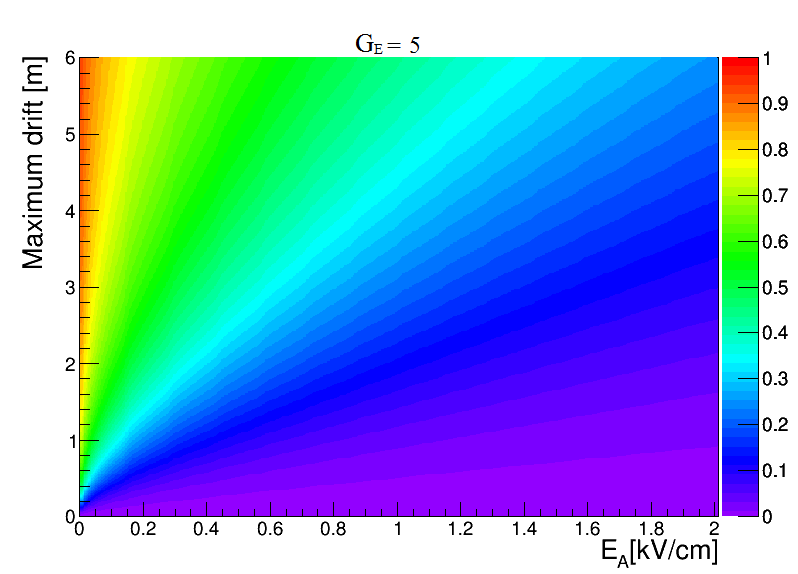}
  \includegraphics[height=.25\textheight]{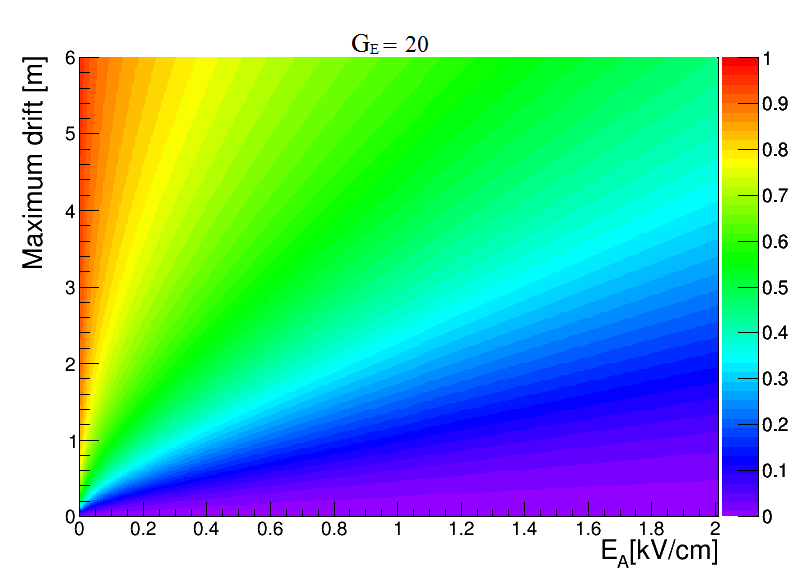}
   \includegraphics[height=.25\textheight]{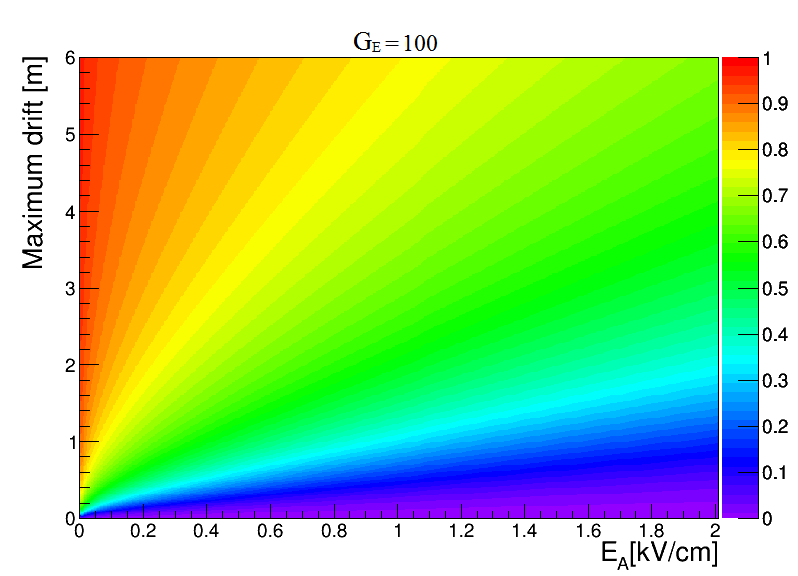}
  \caption{Surface case: recombination probability  (color scale) for electrons generated at the cathode level as a function of the maximum drift length L and the anode field E$_{A}$, considering  $G_I = 0$ ({\it{top-left}}), $G_I = 5$ ({\it{top-right}}),  $G_I = 20$ ({\it{bottom-left}}) and $G_I = 100$ ({\it{bottom-right}}).}
 \label{Losses_Surface_Gfixed}
  \end{center}
\end{figure}

The Eq. \ref{Worst_probability} has been also solved for a detector placed on the surface  considering the same ion gains and field values as the underground case. The corresponding secondary recombination probability  is  plotted in Fig. \ref{Losses_Surface_Gfixed} for a maximum drift length up to 6~m. At a typical field of $\approx$~1~kV/cm,  a measurable value for the recombination ($\approx$~$5\%$) is obtained in case of  a single phase detector  ($G_I = 0$) with maximum drifts of  $2-3$ meters.  We can expect  charge signal losses $\leq$~10\% only for a couple of meters drifts and ion gains of the order of $5-10$, while for $G_I\gtrsim$~20 and L~$\gtrsim$~2~m  at least half of the electrons are expected to recombine with the free ions along the drift. In case of $G_I$  of the order of 100, the largest fraction of the charge created near the cathode is quenched in the LAr target, even at higher fields or smaller drift distances.

We evaluated the signal quenching by the positive charge density in the LAr,  assuming a constant ionization rate with the ions uniformly distributed in the active volume. In order to calculate the electron/ion recombination on an event by event basis, it has to be taken into account that specific areas, with much higher ion current and field distortion compared to the calculated  average values, can be locally produced inside the active volume. While the uniform and constant current approximation is nicely valid for an underground detector, whose ion cloud  is mainly given by  a  large number of \textcolor{black}{submillimeter}  $^{39}$Ar decays,  the ionization paths produced  by the cosmic rays in a shallow detector are nearly vertical tracks  with length comparable to the maximum drift  $L$. In this case, the  ion current will be localized around the muon direction and,  depending on the detector length and the ion velocity, it can last for several minutes after the muon interaction, \textcolor{black}{producing a much larger filed distortion and secondary recombination in that specific region than the one calculated in average. Possibly the detector can be blinded locally until the positive charge cloud is collected by the cathode. }

At the same time, the present discussion has been carried out  considering the liquid argon volume in a steady state, although  the convection motion, given by the temperature gradient inside the detector, and the liquid recirculation, necessary to keep the required argon purity, have to be considered. Given the relatively fast drift time of the electrons ($v_e \approx 2$~mm/$\mu$s \cite{Walkowiak:2000wf}), their drift is not significantly affected by the liquid motion,  however, that could be the case for the   five orders of magnitude slower ions. Only an extremely powerful recirculation flow  (several tens  of  m$^3$/hour) could produce an overall motion barely comparable  with the typical drift speed  of the ions, however, the convection flows, which have been evaluated to be in the range of mm/s \cite{WA105:technical},  could be comparable with the ion drift velocity and, hence, change  their effective drift. \textcolor{black}{In this case, a detailed evaluation of the convection motions within the LAr volume detector has to be carried out in order to assess the effective ion current inside the sensitive region}.

Recent results from the analysis of the  ICARUS data, taken during the commission above ground in 2001, show evidence of the bending of the muon  tracks reconstructed inside the T600 module \cite{ICARUS:2015torti} that is likely given by the space charge induced electric field distortions. The small value of such effect ($\approx$~3~mm maximum) is consistent with the relatively short drift distance (1.5~m) of the detector. Other experiments based on LAr-TPCs, with \textcolor{black}{longer maximum drift lenghts}, showed additional evidences of  space charge effects \cite{Ereditato:2014tya,Mooney:2015kke}, however no indications of a possible charge quenching by electron/ion recombination has been ever reported. 

According to our calculations \textcolor{black}{concerning the surface case, a $\approx~3-4$~\% maximum signal loss  is expected in case of a detector similar to MicroBooNE  ($G_I=0$, $E_d=$~0.27~kV/cm and L~=~2.5~m),  and a factor $\approx~35-40$~\% is obtained considering ARGONTUBE ($G_I=$~0, $E_d=$~0.24~kV/cm and L=~5~m). The quenching given, in average, by the secondary recombination in the former case should be one order of magnitude smaller than the one produced by the measured concentration of electronegative impurities, however the two effects should produce similar losses in the latter experiment.  }   

\begin{figure}[t!]
\begin{center}
\includegraphics[height=.3\textheight]{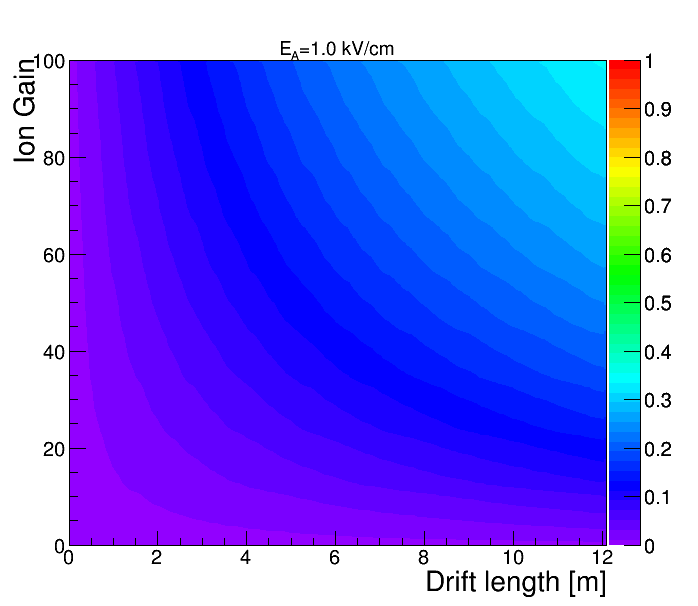}	
  \includegraphics[height=.3\textheight]{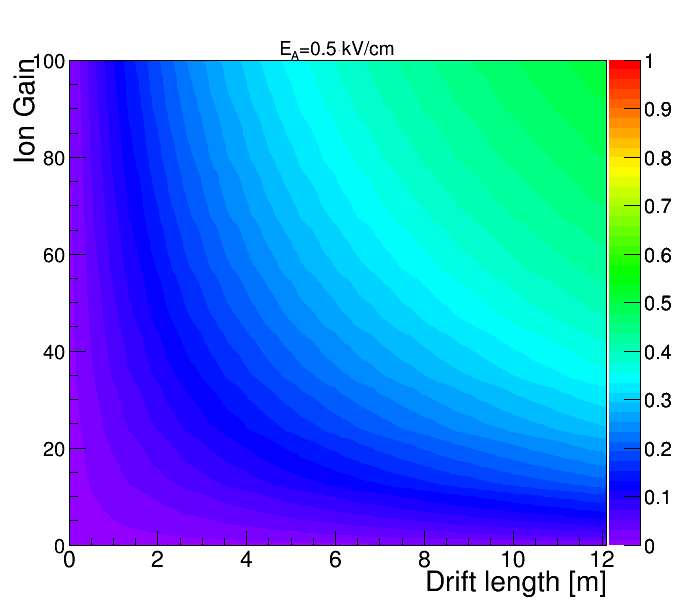}
  \caption{Underground case: recombination probability (color scale) as a function of the ion gain $G_I$ and drift length $l$ for a L = 12 m detector, in case of $E_A$ = 1.0 kV/cm ($left$) and $E_A$ = 0.5 kV/cm ($right$).}
 \label{Losses_UG_Dune}
  \end{center}
\end{figure}

\textcolor{black}{The Deep Underground Neutrino Experiment (DUNE), is especially relevant in this context given its drift length of several meters, and it can provide a clear evidence of the secondary recombination effect produced by the ion current in liquid.} Particularly, the  double phase LAr-TPC, with L = 12~m and charge amplification \cite{Dune:2015}, should be characterized by a measurable electron signal quenching even at small values of $G_I$. In Fig. \ref{Losses_UG_Dune}, the secondary recombination probability is shown as a function of the drift length $l$ for different ion gains, in case of  $E_A$ = 1.0 kV/cm  and $E_A$ = 0.5 kV/cm. In the first case, more than 10~\% of the charge  is expected to recombine in a large section of the detector, even considering relatively small ion amplifications ($G_I\gtrsim 5$). Much larger recombination probabilities, up to  50~\%,  are expected at the lower drift field.  
\textcolor{black}{More detailed calculations concerning the experimental cases will be reported elsewhere \cite{DMCIEMAT}.}
 
The previous results set some limits  on the maximum charge amplification effectively  achievable with a double phase detector operated with natural argon, \textcolor{black}{however the restriction could be  exceeded using radiopure argon, as the one extracted from deep underground reservoirs. Given its scarse availability at the moment, this option is currently impractical  for kilo-ton scale detectors, altough} a radioactive contamination of $^{39}\text{Ar}$ and  $^{85}\text{Kr}$  at a level of $\approx$~10$^{-3}$~Bq/kg, as recently measured by the DarkSide collaboration \cite{DS:2016} \textcolor{black}{A similar activity in a deep underground detector}  could eventually reduce the space charge effect and the recombination probability to a negligible level.

\section{Conclusions} 

The small mobility coefficient  of the positive ions in a liquid argon time projection chamber ensures that they spend considerably longer time in the active volume with respect to the electrons. Measurable space charge effects can be originated by the positive charge  accumulation produced through the muons interactions and the $^{39}\text{Ar}$ decay. In a double phase detector  with an electron signal amplification, that effect can be increased by the injection in the liquid of the ions  produced by the avalanche in the gas phase. We evaluated the impact of the ion current on the uniformity of the electric field, as well as  the charge signal quenching probability due to the drifting electron-ion recombination, as a function of the drift length. According to our calculations, the average signal loss in a single phase underground detector is below 1\% unless very long drifts ($>$~5 m) and relatively low fields ($<$~0.5 kV/cm) are foreseen. 
 Depending on the charge amplification factor in the gas, a double phase underground chamber can be characterized by a relevant field non-uniformity with an electron signal quenching probability larger than 50\%.  Finally the results show a potential concern for the operation of  massive detectors with a maximum drift of many meters and the charge amplification, when operated above the ground.
 To the best of our knowledge, the study evidences, for the the first time, an intrinsic limit for the maximum practical drift obtainable with a TPC operated with natural argon, even in case \textcolor{black}{of a null electronegative impurities concentration}.

\section*{Acknowledgments}
The research has been funded by the Spanish Ministry of Economy and Competitiveness (MINECO) through the grant FPA2015-70657P. The authors were also supported by the ``Unidad de Excelencia Mar\`{i}a de Maeztu: CIEMAT - F\'{i}sica de part\'{i}culas"  through the grant MDM-2015-0509. The authors are also grateful to Dr. Thorsten Lux, Dr. Matteo Cavalli-Sforza, Mr. Manuele Lemme and Mr. Richard Hallett for their inspiring comments and suggestions. 

\begin{appendices}
\section{Calculation of the field lines}
\label{FieldL}

\begin{figure}[!tp]
\begin{center}
  \includegraphics[height=.25\textheight]{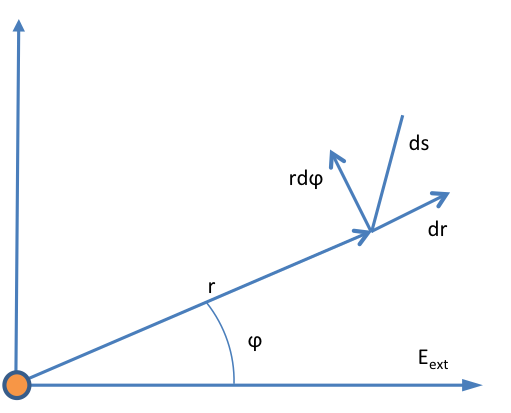}
  \caption{Coordinate system with the ion in the origin and the external electric field in the direction of the $x$ axis.}
  \label{Ion_scheme}
  \end{center}
\end{figure}

We consider a cylindrical symmetry with an arbitrary plane containing the axis and the origin. The section of the flux tubes with the plane gives the field lines, $r(\varphi)$. The coordinate system is defined such that the $x$ axis corresponds to the field direction, and $s$ is the arc length with arbitrary origin (see Fig.~\ref{Ion_scheme}). Due to the cylindrical symmetry, we will only consider the half-plane with positive coordinates.

The external electric field in polar coordinates is expressed as follows:
\begin{equation}
\vec{E}_{ext}=\left( \begin{array}{c}
E_{ext}\,\text{cos}\varphi  \\
-E_{ext}\,\text{sin}\varphi \end{array} \right).
\label{E_Ext}
\end{equation}

The total electric field, which takes into account the external field and the ion field is written as
\begin{equation}
\vec{E}=\vec{E}_{ion}+\vec{E}_{ext}=\frac{q\vec{r}}{4\pi\epsilon r^3}+\vec{E}_{ext}=\left( \begin{array}{c}
\frac{q}{4\pi\epsilon r^2}+E_{ext}\,\text{cos}\varphi  \\
-E_{ext}\,\text{sin}\varphi \end{array} \right).
\label{E_total}
\end{equation}

By definition, the electric field is tangent to any field line, so $\vec{E}=K_0(s)\,\vec{\tau}$, where $\vec{\tau}$ is the unitary vector tangent to a field line in $s$:
\begin{equation}
\vec{\tau}=\frac{\text{d}\vec{r}}{\text{d}s}=\left( \begin{array}{c}
\frac{\text{d}r}{\text{d}s} \\
r\frac{\text{d}\varphi}{\text{d}s}\end{array} \right).
\label{tau}
\end{equation}

Equalising both expressions of the electric field we get:
\begin{eqnarray}
\frac{q}{4\pi\epsilon r^2}+E_{ext}\,\text{cos}\varphi =K_0(s)\frac{\text{d}r}{\text{d}s}.
\label{Dif_eq_1_1}
\end{eqnarray}
\begin{eqnarray}
-E_{ext}\,\text{sin}\varphi=K_0(s)\,r\,\frac{\text{d}\varphi}{\text{d}s}.
\label{Dif_eq_1_2}
\end{eqnarray}

Then, from the Eq.~\ref{Dif_eq_1_2} we isolate d$s$, $\text{d}s=-\frac{K_0(s)r}{E_{ext}\text{sin}\varphi}\text{d}\varphi$, and we substitute this expression in the Eq.~\ref{Dif_eq_1_1} to obtain
\begin{eqnarray}
\frac{q}{4\pi\epsilon r^2}+E_{ext}\,\text{cos}\varphi = -E_{ext}\,\text{sin}\varphi\,\frac{1}{r}\,\frac{\text{d}r}{\text{d}\varphi}.
\label{Dif_eq_2_1}
\end{eqnarray}

We multiply the Eq.~\ref{Dif_eq_2_1} by $r^2/E_{ext}$, getting
\begin{eqnarray}
\frac{q}{4\pi\epsilon E_{ext}}+r^2\,\text{cos}\varphi = -\text{sin}\varphi\,r\,\frac{\text{d}r}{\text{d}\varphi}.
\label{Dif_eq_2_2}
\end{eqnarray}

We perform the change of variable $h=r^2$, $\text{d}h=2r\text{d}r$ in the Eq.~\ref{Dif_eq_2_2} and we also define $K_1=q/4\pi\epsilon E_{ext}$, thus
\begin{eqnarray}
K_1+h\text{cos}\varphi=-\frac{\text{sin}\varphi}{2}\frac{\text{d}h}{\text{d}\varphi},\nonumber\\
\frac{\text{d}h}{\text{d}\varphi}+\frac{2h}{\text{tg}\varphi}=-\frac{2K_1}{\text{sin}\varphi},
\label{Dif_eq_3}
\end{eqnarray}

which is a first order differential equation. The integration factor is:
\begin{equation}
F(\varphi)=\text{exp}\left(\int\frac{2}{\text{tg}x}\text{d}x\right)=\text{exp}\left(\int\frac{2\,\text{cos}x}{\text{sin}x}\text{d}x\right)=\text{exp}(2\,\text{ln}(\text{sin}\varphi))=\text{sin}^2\varphi,
\label{Int_fact}
\end{equation}

and the solution of the Eq.~\ref{Dif_eq_3} is given by
\begin{equation}
h=-\frac{1}{F(\varphi)}\int F(x)\frac{2K_1}{\text{sin}x}\text{d}x=-\frac{2K_1}{\text{sin}^2\varphi}\int \text{sin}x\text{d}x=-\frac{2K_1}{\text{sin}^2\varphi}(-\text{cos}\varphi+C).
\label{Int_h}
\end{equation}

As a result, the expression of the field lines, considering the previous change of variable, $h=r^2$, is the following:
\begin{equation}
r(\varphi)=\sqrt{h}=\sqrt{\frac{-q}{2\pi\epsilon E_{ext}}}\frac{\sqrt{C-\text{cos}\varphi}}{\text{sin}\varphi}.
\label{r_phi}
\end{equation}

The Eq.~\ref{r_phi} defines the field according to the parameter  $C$ (see Fig. \ref{Field_lines}). Considering an external electric field of 1~kV/cm, the value of the scale factor $\sqrt{-q/2\pi\epsilon E_{ext}}$ is 0.14~$\mu$m. Depending on the value of the parameter $C$, we distinguish four different cases:
\begin{enumerate}
\setlength{\itemsep}{0pt}
\setlength{\parskip}{0pt}
\setlength{\parsep}{0pt}
\item If $C>1$, $r(\varphi)$ is defined in the whole interval (0,$\pi$) of $\varphi$. Since $r(\varphi) \rightarrow \infty$ when $\varphi$ approaches to the limits of the interval, the equation represents a line that goes from $-\infty$  to $\infty$.
\item If $-1<C<1$, there are values of $\varphi$ near to 0 for which $r(\varphi)$ is not defined. Since the value of $r(\varphi)$ for the minimum of $\varphi$ is zero, the equation represents a line that goes from $-\infty$  to the origin.
\item If $C<-1$, $r(\varphi)$ is not defined.
\end{enumerate}

The flux tube that includes all the field lines that end in the ion is defined by $C=1$, and the value of the transverse section of the flux tube with $C=1$ in a point far from the ion is:
\begin{equation}
\pi r_y^2(\varphi\rightarrow\pi)=\pi r^2\text{sin}^2\varphi=\pi\frac{-q}{2\pi\epsilon E_{ext}}(1-\text{cos}\varphi)\simeq\frac{-q}{\epsilon E_{ext}}.
\label{lim_sec}
\end{equation}

Fig.~\ref{Field_lines} shows the field lines approaching the ion, positioned at (0,0) which has a negligible size at the micron scale. The lines correspond to different values of $C$, at the same time the red line ($C=1$) gives the envelope of all the field lines ending on the ion.

\section{Cathode voltage calculation}
\label{KVolt}

The cathode voltage necessary to obtain a given field taking into account the ion current  can be calculated  integrating  the drift field expression given by the Eq.~\ref{Drift_field} along the drift path.
\begin{equation}
V(L)=\int_0^L E_d(l)\,\text{d}l=\sqrt{\frac{hq}{\epsilon\mu_i}}\int_0^L\sqrt{l^2+2\frac{\,j_i(0)}{h}\,l+\frac{f_0}{h}}\,\text{d}l.
\label{Drift_field_integral}
\end{equation}

Defining for simplicity $s=l+j_i(0)/h$ and $R=\sqrt{f_0/h-(j_i(0)/h)^2}$ we can get 

\begin{eqnarray}
V(L)&=&\sqrt{\frac{hq}{\epsilon\mu_i}}\int_{j_i(0)/h}^{L+j_i(0)/h}\sqrt{s^2+R^2}\,\text{d}s \nonumber\\
&=&\frac{1}{2}\sqrt{\frac{hq}{\epsilon\mu_i}}\left[s\sqrt{s^2+R^2}+R^2\text{ln}\left(s+\sqrt{s^2+R^2}\right)\right]_{j_i(0)/h}^{L+j_i(0)/h}.
\label{Drift_field_integral_s}
\end{eqnarray}
Substituting the integration limits we finally have
\begin{eqnarray}
V(L)&=&\frac{1}{2}\sqrt{\frac{hq}{\epsilon\mu_i}}\left[\left(L+\frac{j_i(0)}{h}\right)\sqrt{\left(L+\frac{j_i(0)}{h}\right)^2+R^2} 
\right. 
\nonumber\\
&+&
\left. 
R^2\text{ln}\left(L+\frac{j_i(0)}{h}+\sqrt{\left(L+\frac{j_i(0)}{h}\right)^2+R^2}\right)-\frac{j_i(0)}{h}\sqrt{\frac{f_0}{h}}
\right. 
\nonumber\\
&-&
\left. 
R^2\text{ln}\left(\frac{j_i(0)}{h}+\sqrt{\frac{f_0}{h}}\right)\right],
\label{Potential}
\end{eqnarray}
which expresses the cathode voltage as a function of the  constant ionization rate $h$ and the ion current at the anode $j_i(0)$ for a  detector with a given drift length $L$.

\end{appendices}

\end{document}